\documentclass[twocolumn,twoside,journal]{IEEEtran}
\usepackage{epsfig,amsfonts,subfigure,psfrag}
\usepackage[nolist]{acronym}
\usepackage{graphicx,cite,amssymb,amsmath,color}
\usepackage{multicol}
\usepackage{multirow}

\IEEEoverridecommandlockouts

\begin{document}
\begin{acronym}
\acro{LT}{Luby-transform}\acro{CER}{codeworderrorrate}
\acro{ML}{maximum-likelihood} \acro{MPE}{multi-protocol
encapsulation} \acro{FEC}{forward error correction}
\acro{MDS}{maximum distance separable} \acro{RS}{Reed Solomon}
\acro{BEC}{binary erasure channel} \acro{BEEC}{binary
error-and-erasure channel} \acro{w.r.t.}{with respect to}
\acro{BEC}{binary erasure channel}\acro{SEME}{single-error
multiple-erasures} \acro{GE}{Gaussian
elimination}\acro{i.i.d.}{independent and identically distributed}
\acro{LDPC}{low-density parity-check} \acro{BSC}{binary symmetric
channel} \acro{CRC}{cyclic redundancy check} \acro{IT}{iterative}
\acro{GJE}{Gauss-Jordan elimination} \acro{EEC}{error-and-erasure
channel} \acro{PPM}{pulse-position modulation}
\acro{GeIRA}{generalized irregular
repeat-accumulate}\acro{LLR}{log-likelihood ratio}
\acro{LMSC}{land mobile satellite channel}
\acro{HMM}{hidden Markov model}
\acro{BW}{Baum-Welch}
\acro{MAP}{maximum a posteriori}
\acro{LoS}{line of sight}
\acro{p.d.f.}{probability density function}
\acro{STM}{state transition matrix}
\acro{BCJR}{Bahl-Cocke-Jelinek-Raviv}
\acro{SISO}{soft-in soft-out}
\acro{EM}{expectation-maximization}
\acro{SDARS}{satellite digital audio radio service}
\acro{ML}{maximum likelihood}
\acro{GPS}{global positioning system}
\acro{DVB}{digital video broadcasting}
\acro{ESA}{European Space Agency}
\acro{SA}{simulated annealing}
\end{acronym}


\title{On the Application of the Baum-Welch Algorithm for Modeling the Land Mobile Satellite Channel}
\author{
Bal\'azs Matuz, Francisco L\'azaro Blasco and Gianluigi Liva
\thanks{Bal\'azs Matuz, Francisco L\'azaro Blasco and Gianluigi Liva are with the Institute of Communications and
Navigation, German Aerospace Center (DLR), Oberpfaffenhofen, 82234
Wessling, Germany. Email: \{Balazs.Matuz, Francisco.LazaroBlasco,
Gianluigi.Liva\}@dlr.de.}} \maketitle
\thispagestyle{empty} \pagestyle{empty}


\begin{abstract}
Accurate channel models are of high importance for the design of
upcoming mobile satellite systems. Nowadays most of the models for
the \ac{LMSC} are based on Markov chains and rely on measurement
data, rather than on pure theoretical considerations. A key problem
lies in the determination of the model parameters out of the
observed data. In this work we face the issue of state
identification of the underlying Markov model whose model parameters
are a priori unknown. This can be seen as a \ac{HMM} problem. For
finding the \ac{ML} estimates of such model parameters the \ac{BW}
algorithm is \textcolor{black}{adapted} to the context of channel
modeling. Numerical results on \textcolor{black}{test} data
sequences reveal the capabilities of the proposed algorithm. Results
on real measurement data are finally presented.
\end{abstract}


{\pagestyle{plain} \pagenumbering{arabic}}


\section{Introduction}\label{sec:intro}

Satellite services to mobile users are experiencing a renewed
interest thanks to the licenses
granted for S-band usage for broadcast and interactive
services \cite{BOLEAALAMANAC01, SCALISE02, SDARS}. The underlying
communication channel, referred to as \acf{LMSC}, is characterized
by strong variations of the received signal power. Obstacles in the
propagation path between the satellite and the mobile terminal, such
as buildings or trees may cause shadowing or even a complete
blockage of the signal. With increasing frequency and decreasing
elevation angle such events become more and more likely and strongly
impact service availability. A further source of fading is due to
multipath propagation: objects in the vicinity of the receiver are
source of reflections that cause constructive or destructive
interference. 
In the past several authors proposed Markov chain models to describe
the behavior of the \ac{LMSC} \cite{LUTZ91,
KARASAWA97, PEREZFONTAN01, ALASSEUR08, SCALISE01}. The modeling
approach can be divided into two stages. First a Markov chain is set
up to model slow transitions between different signal levels due to
\textcolor{black}{macroscopic effects such as} blockage, shadowing,
etc. In practice models with two or three states are common, but
also a larger number of states is possible. Second, fast signal
variations within each state \textcolor{black}{due to multipath} are
taken into account assuming that the signal amplitude follows some
specific distribution. To give an example, a Ricean distribution may
be used to describe the signal amplitude in \ac{LoS} conditions,
whereas the amplitude in a blockage state could be assumed to be
Rayleigh distributed.

Knowing the underlying channel model, a major issue consists in how
to determine the model parameters out of a sequence of measurement
data. In literature there exist several approaches, most of them
being rather simple and empirical. In \cite{PEREZFONTAN01} the
authors propose first to associate with each measurement sample a
state of the underlying Markov chain. This association is done
manually. Then, for each state the distribution of the associated
samples is approximated by some known distribution by means of curve
fitting. In \cite{LUTZ91} the weighted sum of some known
distributions is fitted to the \ac{p.d.f.} of the measured data.
This gives the parameters for the distributions in the different
states. Then, each sample is associated with a state by placing
thresholds on the signal level. The thresholds are put according to
the state probabilities from the fitting step. For highly
overlapping distributions, this only works with limited accuracy, as
we will show later. A more rigorous attempt is the technique in
\cite{ALASSEUR08} based on reversible jump Monte Carlo computation
\cite{GREEN95}. It suggests fully blind estimation, making no prior
assumptions on the number of states, nor on the specific
distributions, allowing huge flexibility. However, this has the
price of a significant increase in complexity and the resulting
states and distributions often lack \textcolor{black}{sufficient}
explanations in terms of underlying physical effects.

Within this work we propose a further way to estimate the model
parameters. It exploits the fact that the state identification can
be seen as a \acf{HMM} problem: out of the channel observation we
would like to draw conclusions about the underlying Markov process
that is not directly observable. A solution to this problem is given
by the \acf{BW} algorithm, that has been widely used in other
fields, such as speech or pattern recognition. An application to
models of digital channels has already been provided in
\cite{Turin93:ModelingError}. In the sequel our focus is on the
\ac{LMSC}. We impose some constraints on the \ac{BW} algorithm in
order to improve its convergence and for sake of simplification. In
particular it is well-known that its convergence properties depend
on the initial model assumptions. Hence, unlike in \cite{ALASSEUR08}
we assume prior knowledge on the type of distributions and the
distribution parameters (to be obtained by a preceding curve fitting
step). Also, we fix the number of states in advance.


The remaining part of the paper is organized as follows. In Section
\ref{sec:BW_overview} we recap the \ac{BW} algorithm. Further we introduce
 a log-domain computation of the forward-backward metric of the \ac{BW}
 algorithm and discuss some adaptations. Section \ref{sec:BW_application}
  reports the performance of the algorithm on \textcolor{black}{test} data,
  as well as on \textcolor{black}{sequences of real} measurement data. A comparison with the method in
  \cite{LUTZ91} is provided. 

\section{Overview of the \ac{BW} Algorithm}\label{sec:BW_overview}

Following the footsteps of \cite{RABINER89}, we consider next the
problem of  associating an observed sample with a state of our
\ac{HMM}. The \ac{BW} algorithm can be applied to maximize the
probability of a state given the entire observation sequence.
\textcolor{black}{Let us denote as $X_t$ the state of the \ac{HMM}
at time $t$, and as $\mathbf{r}=(r_1, r_2, \ldots, r_n)$ the vector
of $n$ observations. The problem can be formalized as follows: given
the vector of $\mathbf{r}$, we are interested in locally calculating
the \textcolor{black}{probability} of being in state $i$ at time
$t$, i.e. $\Pr \{X_t=i | \mathbf{r} \}$.} We define
\begin{equation}
g_t(i)\triangleq \Pr \{X_t=i | \mathbf{r} \}= \frac{f_{\mathbf{R},X_t}(\mathbf{r},X_t=i)}{f_\mathbf{R}(\mathbf{r})} .\label{eq:APP_bayes}
\end{equation}
\textcolor{black}{For now, we focus on the joint \ac{p.d.f.} in the
enumerator. Under the \ac{HMM} assumption, after dropping the
subscripts for simplicity,
 this \ac{p.d.f.} can 
be rewritten as}
\begin{equation}
\begin{array}{cccc}
{f(\mathbf{r},X_t=i)}= & \underbrace{f(\mathbf{r}_1^t,X_t=i)} & \cdot & \underbrace{f(\mathbf{r}_{t+1}^n|X_t=i)}\\
 & a_t(i)& & b_t(i) \label{eq:split}
\end{array}.
\end{equation}
Here we used the shorthand $\mathbf{r}_k^w$ to  denote the elements
$(r_k, r_{k+1}, \ldots, r_w)$ of the  observation sequence
$\mathbf{r}$, with $w>k$. Further, referring to \eqref{eq:split}, we
define a forward metric $a_t(i)$ and a backward metric $b_t(i)$.
It follows that
\begin{equation}
g_t(i)=\frac{a_t(i)b_t(i)}{\sum_{i=1}^m a_t(i)b_t(i)}, \label{eq:def_g}
\end{equation}
where the normalization by $\sum_{i=1}^m a_t(i)b_t(i)$ corresponds to $f_\mathbf{R}(\mathbf{r})$ in \eqref{eq:APP_bayes} and $m$ denotes the number of states. Being $p_{ij}$ the transition probability from state $i$ to $j$, $p_i$ the probability of state $i$ and $f_i(r)$ the probability density function given state $i$, the forward and the backward metric can be computed iteratively as
\begin{equation}
a_t(i)=f_i(r_t)\sum_{j=1}^m a_{t-1}(j)p_{ji},\label{eq:def_a}
\end{equation}
\begin{equation}
b_t(i)=\sum_{j=1}^m b_{t+1}(j)p_{ij}f_j(r_{t+1}),\label{eq:def_b}
\end{equation}
with the initial metrics $a_1(i)=p_i f_i(r_1)$ and $b_n(i)=1 \forall i$.
\begin{figure}
  \begin{center}
   \includegraphics[width=0.95\columnwidth]{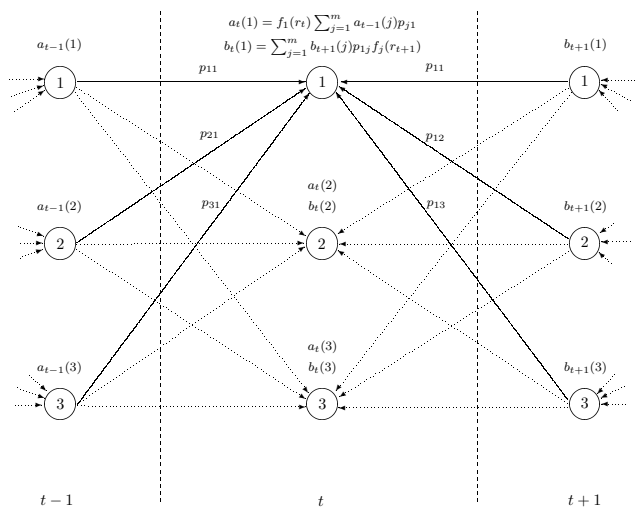}\\
  \caption{Calculation of the forward-backward metric for 3 states (linear domain).}\label{fig:Prob_Domain_FB}
  \end{center}
\end{figure}

Figure \ref{fig:Prob_Domain_FB} shows an excerpt of a state diagram for a Markov chain in the interval $[t-1, t+1]$. The nodes of the trellis diagram at each time instance denote one of the three possible states, whereas the lines denote all possible transitions. Consider for example state $1$. As indicated by the solid arrows, the metrics from all the nodes at $t-1$, as well as $t+1$ contribute to the calculation of the probability of state $1$ at time $t$. The most likely state sequence can be determined by choosing the state with the highest probability at each time instance.

Further, we may wish to estimate the probability of having a transition from state $i$ at time $t$ to state $j$ at time $t+1$, given the observation $\mathbf{r}$. This can be expressed as
\[
z_t(i,j)\triangleq \Pr \{X_t=i,X_{t+1}=j|\mathbf{r}\},
\]
and it turns out that
\begin{equation}
z_t(i,j)=\frac{a_t(i)p_{ij}f_j(r_{t+1})b_{t+1}(j)}{\sum_{i=1}^m
{\sum_{j=1}^m a_t(i)p_{ij}f_j(r_{t+1})b_{t+1}(j)}}.
\label{eq:transition_probability_at_time_t}
\end{equation}
Even if the \ac{BW} algorithm is in principle more general, we
restrict ourselves to the simple case where the density functions $f_i(r)$,
$i=1,\ldots, m$, are perfectly known, whereas we do not have any
knowledge about the transition probabilities $p_{ij}$, 
which we want to estimate. To do so, we chose some initial values for $p_{ij}$,\footnote{In principle the choice is arbitrary. Nevertheless initial values not too far from the real values facilitate the convergence of the \ac{BW} algorithm. Good starting points can be found in literature (e.g. in \cite{LUTZ91}).}  
 run the forward-backward algorithm and re-estimate the transition
probabilities $\hat{p}_{ij}$ and the initial state probabilities $\hat{p}_i$ according to the re-estimation formulae
\[
\hat{p}_{ij}=\frac{\sum_{t=1}^{n-1}z_{t}(i,j)}{\sum_{t=1}^{n-1}\sum_{j=1}^m
z_{t}(i,j)},\qquad \hat{p_i}=\sum_{j=1}^m z_1(i,j).
\]
Former values of $p_{ij}$ and $p_i$ are replaced by the new
estimates
 and the forward-backward algorithm is run again, leading to updated estimates, which are then fed-back. This process is iterated several times. The state identification step corresponds to the E-step of the \ac{EM} algorithm, where the model parameters are assumed to be fixed. The re-estimation step corresponds to the M-step, where the most likely model parameters are determined given the hidden state sequence.

A final remark is related to the convergence of the algorithm. It is well known that the \ac{EM} algorithm, as well as its special instance, the \ac{BW} algorithm, increases the likelihood of the model iteration by iteration till it converges to a maximum value \cite{TURIN04}. However, the algorithm may converge to a local maximum of the likelihood function, rather than to a global one. The convergence of the algorithm can be facilitated by limiting the set of a priori unknown model parameters. Alternatively, a set of various starting points can be considered.

\subsection{Log-domain Implementation of the \ac{BW} Algorithm}\label{sec:log_domain}
Already for short observation sequences ($n>100$) the forward-backward metric may get numerically unstable. As a solution, for each $t$ a normalization of the metric is usually performed \cite{TURIN04}. Alternatively, a log-domain representation of the corresponding equations is proposed here.
Let us define log-probabilities as $\gamma_t(i)\triangleq \ln g_t(i)$,
$\alpha_t(i)\triangleq \ln a_t(i)$ and $\beta_t(i)\triangleq \ln
b_t(i)$, with $\ln(\cdot)$ being the natural logarithm. Then, \eqref{eq:def_g} can be rewritten as
\[
\begin{array}{cc}
\gamma_i(t)=\alpha_i(t)+\beta_i(t)-\ln \sum_{i=1}^{m} & \underbrace{\exp (\alpha_i(t)+\beta_i(t))}\\
 & \exp(\kappa_i) \\
\end{array}.\label{eq:gamma_semilog}
\]
Note that the last term can be solved in the log-domain by applying recursively the so-called $\max  ^*$ operator (also known as Jacobi logarithm) that is defined as $\max^*(\kappa_1,\kappa_{2})\triangleq \ln\left(\exp(\kappa_1)+\exp(\kappa_{2})\right)$.  Exploiting the identity
\[
{\max}^*(\kappa_1,\kappa_{2})=\max(\kappa_1,\kappa_{2})+\ln\left(1+\exp(-|\kappa_1-\kappa_{2}|)\right) \label{eq:max_star}
\]
and noticing that ${\max}^*(\kappa_1,\kappa_{2},\kappa_{3})$ can be
recursively calculated as
${\max}^*(\kappa_{1},{\max}^*(\kappa_{2},\kappa_{3}))$, we have
\[
\gamma_i(t)=\alpha_i(t)+\beta_i(t)- {\max_{i=1:m}}^*
(\alpha_i(t)+\beta_i(t)). \label{eq:gamma_log}
\]
In a similar manner, using the shorthand $\phi_i(r_t)\triangleq \ln f_i(r_t)$, $\pi_{ij}\triangleq \ln p_{ij}$ and $\pi_{i}\triangleq \ln p_{i}$ we have that the recursions
\[
\alpha_i(t)=\phi_i(r_t) + {\max_{j=1:m}}^*(\alpha_j(t-1)+\pi_{ji})
\]
with $\alpha_i(1)=\pi_i+\phi_i(r_1)$ and
\[
\beta_i(t)={\max_{j=1:m}}^*(\beta_j(t+1)+\pi_{ij}+\phi_j(r_{t+1}))
\]
with $\beta_i(n)=0$, $\forall i$. Finally, for the re-estimation of the \ac{BW} metrics we define $\zeta_t(i,j)\triangleq \ln z_t(i,j)$. Taking  \eqref{eq:transition_probability_at_time_t} we have
\[
\zeta_{t}(i,j)= \alpha_t(i)+\pi_{ij}+\phi_j(r_{t+1})+\beta_{t+1}(j)-  \hspace{30 mm}
\]
\vspace{-5 mm}
\[
-{\max_{i=1:m}}^*\left({\max_{j=1:m}}^*\left(\alpha_t(i)+\pi_{ij}+\phi_j(r_{t+1})+\beta_{t+1}(j)\right)\right).
\]
The estimation of the parameters proceeds as
\[
{\pi}_{ij}={\max_{t=1:n-1}}^*\zeta_{t}(i,j)-{\max_{t=1:n-1}}^*\left({\max_{j=1:m}}^*\zeta_{t}(i,j)\right)
\]
while
\[
\pi_i={\max_{j=1:m}}^*\zeta_{1}(i,j).
\]

\subsection{Restrictions on the \ac{BW} Algorithm}
\textcolor{black}{For modeling the \ac{LMSC} applying the \ac{BW}
algorithm some a priori restrictions on the channel parameters have
been applied}. This is mainly motivated by two reasons. First, if
reasonably good estimates of some channel parameters are available,
their use may facilitate the convergence of the algorithm. Second,
we consider important that the obtained results have a
\textcolor{black}{clear physical interpretation}. To give an
example, we would like states to be associated with different
physical events, such as blockage of the signal or direct \ac{LoS}.
During this work \textcolor{black}{ the following restrictions have
been applied}:
\begin{itemize}
  \item \textcolor{black}{The type of distributions to be used has been fixed in advance}. The original \ac{BW} algorithm allows estimating the densities $f_i(r)$ iteratively as a mixture of Gaussian distributions \cite{RABINER89}. It is however well-established that typical propagation conditions (blockage or \ac{LoS}, for instance) can be accurately modeled by known distributions.
  \item \textcolor{black}{Estimates of the distribution parameters are provided to the \ac{BW} algorithm}. Such estimates can be obtained for instance by a curve fitting step and are kept fixed through the \ac{BW} re-estimation. Alternatively at each iteration the estimates could be refined, given the intermediate results.
  \item \textcolor{black}{The number of states is fixed in advance corresponding to some physical events, such as total blockage of the signal by obstacles or \ac{LoS} }.
\end{itemize}

\section{Applications of the \ac{BW} Algorithm}\label{sec:BW_application}
The capabilities of the \ac{BW} algorithm on different data sets are
evaluated next. First we generate artificially a
\textcolor{black}{test} sequence of samples and run iterative
re-estimation. Knowing the original model parameters, our goal is to
assess the quality of the re-estimations provided by the \ac{BW}
algorithm. A comparison with the commonly used threshold method and
some derivatives is done. Second, the \ac{BW} algorithm is applied
to data obtained from a measurement campaign.

\subsection{Application on \textcolor{black}{Test Data Sequences}}
Given a Markov chain with transition probabilities $p_{ij}$ we
generate a sequence of states $\mathbf{x}=(x_1, x_2, \ldots, x_n)$.
For each state, an observation sample according to the associated
\ac{p.d.f.} is produced.  For simplicity, we fix the number of
states $m$ to 2. For state 1, we choose a Gaussian distribution with
standard deviation $\sigma_1=0.2$. To perform different tests, the
mean value $\mu_1$ ranges from $0.4$ to $0.9$. The Gaussian
distribution associated with state 2 has mean $\mu_2 = 1$ and
variance $\sigma_2=0.2$. Since typically the \ac{LMSC} is highly
correlated \cite{PEREZFONTAN01}, we choose the state transition
probabilities of the Markov chain
\[
\begin{array}{ccc}
\left[
  \begin{array}{cc}
    p_{11} & p_{12}\\
    p_{21} & p_{22}\\
  \end{array}
\right]& = &
\left[
  \begin{array}{cc}
    0.950 & 0.050\\
    0.025 & 0.975\\
  \end{array}
\right]
 \end{array},
\]
with corresponding state probabilities $p_1=0.333$ and $p_2=0.667$.
The length of the state sequence (observation sequence) was set to
$n=100000$.

\begin{figure}
  \begin{center}
   \includegraphics[width=0.90\columnwidth]{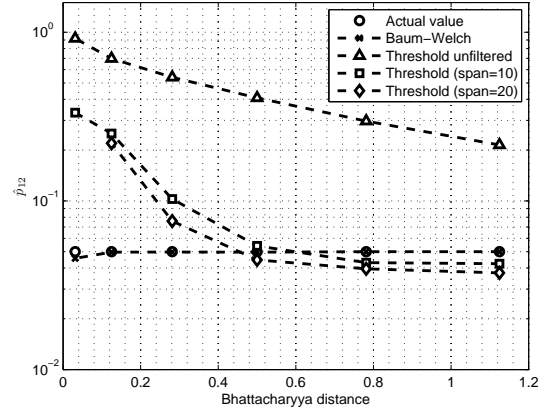}
  \end{center} \caption{Re-estimated transition probability $\hat{p}_{12}$ vs. Bhattacharyya distance for the \ac{BW} algorithm and threshold methods.}  \label{fig:p12_vs_BAtest}
\end{figure}

Given the observations $\mathbf{r}=(r_1, r_2, \ldots, r_n)$ and the
knowledge on the \acp{p.d.f.}, our iterative re-estimation algorithm
is ran to determine the state sequence $x_t$, for $t=1\ldots n$, as well as the
transition probabilities $p_{ij}$ and the state probabilities $p_i$.
It should be obvious that the closer the mean values of both
Gaussian distributions are, the bigger shall be the deviation
between the re-estimated state sequence
$\mathbf{\hat{x}}=(\hat{x}_1, \hat{x}_2, \ldots, \hat{x}_n)$, the associated re-estimated
transition probabilities $\hat{p}_{ij}$, as well as state probabilities
$\hat{p}_i$ and the actual values. To measure the
distance of the two distributions $f_1(r), f_2(r)$ associated with
the two states, we make use of the Bhattacharyya distance
\[
\mathcal{B}\left(f_1(r),f_2(r)\right) = -\ln \int_{-\infty}^{\infty} \sqrt{f_1(r) \cdot f_2(r)}   \; dr .
\]

For sake of comparison we also apply the threshold method to
separate the states \cite{LUTZ91}. Samples below the threshold
$\tau$ are associated with one state, the ones above with the other.
We select the threshold $\tau$, such that the average error
probability
\[
\overline{p}_e \triangleq p_1 \int_{\tau}^{\infty} f_1(r) \; dr + p_2 \int_{-\infty}^{\tau} f_2(r) \; dr,
\]
is minimized. In addition, we assume a priori knowledge of the state probabilities (which could be e.g. provided by a previous curve-fitting step). For two Gaussian distributions with variances $\sigma_1=\sigma_2=\sigma$, this yields
\[\label{eq:threshold}
\tau=\frac{\mu_1+\mu_2}{2}+\frac{\sigma^2 \ln\frac{p_1}{p_2}}{(\mu_2-\mu_1)}.
\]
Further, to suppress frequent state transitions (crossing of the
threshold) we apply moving average filtering on the observation
sequence. The span of the moving average is set to 10 or 20
samples.

\begin{figure}
  \begin{center}
   \includegraphics[width=0.90\columnwidth]{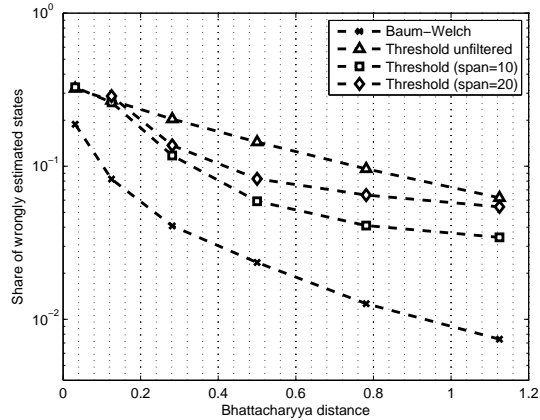}
  \end{center} \caption{Share of wrongly identified states vs. Bhattacharyya distance for the \ac{BW} algorithm and threshold methods. A state at time $t$ is considered to be wrongly identified if $\hat{x}_t \neq x_t$.}  \label{fig:errors_vs_BAtest}
\end{figure}

Figure \ref{fig:p12_vs_BAtest} illustrates the estimated
transition probability $\hat{p}_{12}$ versus the Bhattacharyya
distance for the \ac{BW} algorithm and the threshold methods with and w/o filtering. Despite close mean values of both distributions, the \ac{BW} algorithm provides always accurate estimates for the state transition probability $p_{12}$, whereas the
threshold methods typically fail when $\mathcal{B}(f_1(r),f_2(r))<0.4$. Empirically we found
that best results for the threshold methods can be obtained with an
averaging window span of 10 samples. It shall be noted however that
in this case the resulting state probabilities deviate remarkably as
illustrated in Table \ref{tab:p1}. It turns out that with increasing averaging window size even for $\mathcal{B}(f_1(r),f_2(r))=0.78$ the estimated state probability $\hat{p}_1$ is too low. Figure \ref{fig:errors_vs_BAtest} depicts the share of wrongly labeled states in the estimated state sequence $\mathbf{\hat{x}}$, obtained through a comparison of the original state sequence $\mathbf{x}$ with $\mathbf{\hat{x}}$. Again the \ac{BW} algorithm provides by far the best results, followed by the threshold methods with filtering. Note that for low Bhattacharyya distances the share of errors converges to 0.33 which corresponds to $p_1$.

\begin{table}
  \caption{$\hat{p}_1$ for \ac{BW} and various threshold methods: unfiltered (T1),  filtered with window size 10 (T10) and 20 (T20) samples.}\label{tab:p1}
  \begin{center}
    \begin{tabular}{|c|c|c|c|c|c|}
    \hline
       {\bf $\mathcal{B}$} & { $\mu_2-\mu_1$} &   { BW} &   { T1} &  { T10} &  { T20} \\
    \hline
    \hline
          1.13 &       0.60 &       0.33 &       0.33 &       0.31 &       0.29 \\
    \hline
          0.78 &       0.50 &       0.33 &       0.32 &       0.30 &       0.28 \\
    \hline
          0.50 &       0.40 &       0.33 &       0.31 &       0.28 &       0.26 \\
    \hline
          0.28 &       0.30 &       0.33 &       0.28 &       0.22 &       0.20 \\
    \hline
          0.13 &       0.20 &       0.33 &       0.22 &       0.07 &       0.04 \\
    \hline
          0.03 &       0.10 &       0.33 &       0.08 &       0.00 &       0.00 \\
    \hline
    \end{tabular}
    \end{center}
\end{table}

\subsection{Application on Measurement Data}\label{sec:artificial_data}
In fall 2008 a vast measurement campaign was carried out along the
US East Coast in the framework of the \ac{ESA} funded MiLADY project
\cite{MiLADY}. During the field trials the signal levels of the four
\ac{SDARS} satellites were recorded with a mobile vehicular
receiver. A statistical channel model was derived out of the
collected measurement data employing the \ac{BW} algorithm.
The proceedings are as follows: we first perform a curve-fitting
step on the overall \ac{p.d.f.} of $\mathbf{r}$ similar to
\cite{LUTZ91}. We obtain parameters of the distributions in the
different states which serve as input for the \ac{BW} algorithm. The
resulting state probabilities are used to initialize $a_1(t)$. The
curve fitting is performed using \ac{SA} \cite{SimuAnn}, a fast
meta-heuristic method for global optimization. In case the function
to be optimized has several local maxima \ac{SA} may overcome these
and converge to the global minimum. Following literature, we chose
three simple distributions to characterize the fast signal
variations in the different states. The signal amplitude is assumed
to follow a Rice distribution in case of direct \ac{LoS} to the
satellite. We associate a lognormal distribution with the shadowing
state and assume that the signal amplitude in the blockage state is
Rayleigh distributed.
As second step, a preprocessing stage is required. The fast signal variations within a state are known to be correlated (see e.g. \cite{JAKES}). 
However, \eqref{eq:split} implicitly assumes independency among
observation samples given a certain state. To comply with the
independence assumption, the measurement data is down-sampled,
taking into account the coherence time of the process\footnote{The
\textcolor{black}{spatial separation} between samples after
down-sampling was chosen to be $1~m$ in accordance with
\cite{LUTZ91}. }. This leads the final observation $\mathbf{r}$.
Finally, given the three distributions and the observed sequence,
the \ac{BW} algorithm is applied as described in Section
\ref{sec:BW_overview}.



\begin{figure}
  \begin{center}
   \includegraphics[width=0.90\columnwidth]{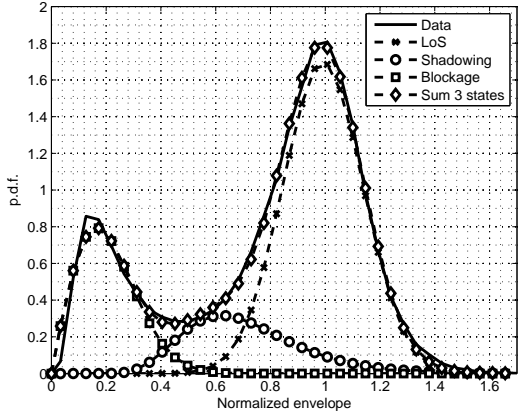}
  \caption{Curve fit on measurement data for urban environment and a satellite elevation of $30^\circ$.}
  \label{fig:curve_fit}
  \end{center}
\end{figure}

Let's consider a typical US urban environment with a satellite
elevation of $30^\circ$. The solid line in Figure
\ref{fig:curve_fit} shows the \ac{p.d.f.} of the measured signal
envelope, whereas the dashed lines with markers give the results of
curve fitting using the three \acp{p.d.f.} specified previously. The
weighted sum of the Rice, lognormal and Rayleigh distributions is
also plotted (dashed with diamonds) and turns out to be close to the
\ac{p.d.f.} of the measured data. The Bhattacharyya distance between
the lognormal (Rayleigh) and Rice (lognormal) distribution is 0.5
(1.1), thus posing challenges for state identification. Table
\ref{tab:D} gives the mean state durations $\bar{D}_i=1/(1-p_{ii})$
and state probabilities $p_{i}$ obtained with different state
identification methods with minimum state duration set to $1~m$.
Results for the \ac{BW} algorithm, the threshold method from
\cite{LUTZ91}, as well as the modified threshold method with a
filter length of 10 samples are shown. It can be observed that the
\ac{BW} algorithm and the threshold method yield the same state
probabilities $\hat{p}_i$ as obtained by means of curve fitting.
However, the mean state durations obtained by the threshold method
are very short. As illustrated in Figure \ref{fig:p12_vs_BAtest} the
threshold method tends to over-dimension $\hat{p}_{ij}$, thus to
under-dimension $\bar{D}_i$. If a prior filtering step is applied,
the state durations become longer than the ones obtained with
\ac{BW}. This is caused by an under-dimensioning of $\hat{p}_{ij}$
for Bhattacharyya distances greater than 0.5 (c.f. Figure
\ref{fig:p12_vs_BAtest}). Here, the state probabilities are no
longer preserved.

\begin {table}
\caption{Mean state duration $\bar{D}$ in meters and state
probability $p_{i}$ for the $3$ propagation states.}\label{tab:D}
\begin{center}
\begin{tabular}{|r|c|c|c|c|}
\hline
           &     Method &        LoS &  Shadowing &   Blockage \\
\hline
\hline
   \multirow{3}{*}{$p_{i}$} &         BW &       0.66 &       0.14 &       0.20 \\

  & T1 &       0.66 &       0.14 &       0.20 \\

          & T10 &       0.70 &       0.12 &       0.18 \\
\hline
 \multirow{3}{*}{$\bar{D}_i$} &         BW &      22.42 &       4.11 &      26.88 \\

           &  T1 &       7.25 &       1.48 &       8.90 \\

           & T10 &      64.35 &       6.87 &      25.15 \\
\hline
\end{tabular}
\end{center}
\end{table}

\section{Conclusion}\label{sec:conclusion}
This work investigates the application of the \ac{BW} algorithm to determine the
parameters for a \ac{LMSC} model out of a set of measurement data.
The \ac{BW} algorithm, allows estimating iteratively the hidden state sequence and the transition
probabilities of the underlying \ac{HMM} even for highly overlapping
states. Especially in environments with frequent shadowing events conventional methods, such as the threshold methods may lead to inaccurate results on the \ac{STM} of the hidden Markov process.
Adaptations of the \ac{BW} algorithm presented here guarantee numerical
stability, as well as proper convergence at manageable complexity.
Adaptations to channels different from the \ac{LMSC} are possible
and may be a matter of future investigation.

\section{Acknowledgements}\label{sec:acknowledgements}
This work was \textcolor{black}{partly} carried out in the framework
of the ESA funded MiLADY project (contract no. 21159/07/NL/GLC).
\textcolor{black}{The authors would like to acknowledge Dr. S.
Scalise and R. Prieto Cerdeira for the useful discussions.}
\bibliography{IEEEabrv,total}

\begin{thebibliography}{10}
\providecommand{\url}[1]{#1}
\csname url@samestyle\endcsname
\providecommand{\newblock}{\relax}
\providecommand{\bibinfo}[2]{#2}
\providecommand{\BIBentrySTDinterwordspacing}{\spaceskip=0pt\relax}
\providecommand{\BIBentryALTinterwordstretchfactor}{4}
\providecommand{\BIBentryALTinterwordspacing}{\spaceskip=\fontdimen2\font plus
\BIBentryALTinterwordstretchfactor\fontdimen3\font minus
  \fontdimen4\font\relax}
\providecommand{\BIBforeignlanguage}[2]{{%
\expandafter\ifx\csname l@#1\endcsname\relax
\typeout{** WARNING: IEEEtran.bst: No hyphenation pattern has been}%
\typeout{** loaded for the language `#1'. Using the pattern for}%
\typeout{** the default language instead.}%
\else
\language=\csname l@#1\endcsname
\fi
#2}}
\providecommand{\BIBdecl}{\relax}
\BIBdecl

\bibitem{BOLEAALAMANAC01}
\BIBentryALTinterwordspacing
A.~Bolea~Alamanac, P.~Burzigotti, R.~De~Gaudenzi, G.~Liva, H.~N. Pham, and
  S.~Scalise, ``In-depth analysis of the satellite component of {DVB-SH}:
  Scenarios, system dimensioning, simulations and field trial results,''
  \emph{International Journal of Satellite Communications and Networking},
  vol.~27, no. 4-5, pp. 215--240, 2009. [Online]. Available:
  \url{http://dx.doi.org/10.1002/sat.933}
\BIBentrySTDinterwordspacing

\bibitem{SCALISE02}
S.~Scalise, C.~Niebla, G.~Gallinaro, M.~Andrenacci, R.~Rinaldo, O.~Del
  Rio~Herrero, M.~Breiling, D.~Finocchiaro, J.~Cebrian~Puyuelo, and
  G.~Schlueter, ``System design for {Pan-European} {MSS} services in
  {S}-band,'' in \emph{Advanced Satellite Multimedia Systems Conference (ASMS)
  and the 11th Signal Processing for Space Communications Workshop (SPSC), 2010
  5th}, 2010, pp. 538 --545.

\bibitem{SDARS}
\BIBentryALTinterwordspacing
Satellite Digital Audio Radio Service (SDARS). [Online]. Available:
  \url{http://www.sirius.com/}
\BIBentrySTDinterwordspacing

\bibitem{LUTZ91}
E.~Lutz, D.~Cygan, M.~Dippold, F.~Dolainsky, and W.~Papke, ``The land mobile
  satellite communication channel - recording, statistics, and channel model,''
  in \emph{IEEE Trans. Vehicular Technology}, vol.~40, no.~2, May 1991, pp.
  375--386.

\bibitem{KARASAWA97}
Y.~Karasawa, K.~Kimura, and K.~Minamisono, ``Analysis of availability
  improvement in {LMSS} by means of satellite diversity based on three-state
  propagation channel model,'' in \emph{IEEE Trans. Vehicular Technology},
  vol.~46, no.~4, November 1997, pp. 957--1000.

\bibitem{PEREZFONTAN01}
F.~Perez-Fontan, M.~Vazquez-Castro, C.~Cabado, J.~Garcia, and E.~Kubista,
  ``Statistical modeling of the lms channel,'' in \emph{IEEE Trans. Vehicular
  Technology}, vol.~50, no.~6, November 2001, pp. 1549--1567.

\bibitem{ALASSEUR08}
C.~Alasseur, S.~Scalise, L.~Husson, and H.~Ernst, ``A novel approach to model
  the land mobile satellite channel through reversible jump {Markov} chain
  {Monte Carlo} technique,'' \emph{IEEE Transactions on Wireless
  Communications}, vol.~7, pp. 532--542, 2008.

\bibitem{SCALISE01}
S.~Scalise, H.~Ernst, and G.~Harles, ``Measurement and modeling of the land
  mobile satellite channel at {Ku}-band,'' \emph{Vehicular Technology, IEEE
  Transactions on}, vol.~57, no.~2, pp. 693 --703, 2008.

\bibitem{GREEN95}
P.~Green, ``Reversible jump {MCMC} computation and {Bayesian} model
  determination,'' in \emph{Biometrika}, vol.~82, no.~40, 1995, pp. 711--732.

\bibitem{Turin93:ModelingError}
W.~Turin and M.~M. Sondhi, ``{Modeling Error Sources in Digital Channels},''
  \emph{IEEE Journal on Selected Areas in Communications}, vol.~11, pp.
  340--347, 1993.

\bibitem{RABINER89}
L.~R. Rabiner, ``A tutorial on hidden {Markov} models and selected applications
  in speech recognition,'' in \emph{Proceedings of the IEEE}, February 1989,
  pp. 77(1):257--286.

\bibitem{TURIN04}
W.~Turin, \emph{Performance Analysis and Modeling of Digital Transmission
  Systems (Information Technology: Transmission, Processing and Storage)},
  1st~ed.\hskip 1em plus 0.5em minus 0.4em\relax Springer, May 2004.

\bibitem{MiLADY}
\BIBentryALTinterwordspacing
Mobile satellite channeL with Angle DiversitY. [Online]. Available:
  \url{http://telecom.esa.int/telecom/www/object/index.cfm?fobjectid=29020}
\BIBentrySTDinterwordspacing

\bibitem{SimuAnn}
S.~Kirkpatrick, C.~D. Gelatt, and M.~P. Vecchi, ``{Optimization by Simulated
  Annealing},'' \emph{Science, Number 4598, 13 May 1983}, vol. 220, 4598, pp.
  671--680, 1983.

\bibitem{JAKES}
W.~C. Jakes, \emph{Microwave mobile communications}.\hskip 1em plus 0.5em minus
  0.4em\relax Wiley, New York, 1974.

\end{thebibliography}

\end{document}